# Bubbles in sheared two-dimensional foams.

C. Quilliet[1,*], M. A. P. Idiart[2], B. Dollet[1], L. Berthier[1], A. Yekini[1]

[1] *Laboratoire de Spectrométrie Physique, B.P. 97,
F-38402 Saint-Martin d'Hères Cedex, France.*

[2] *Instituto de Física, Universidade Federal do Rio Grande do Sul, C.P. 15051,
991501-970 Porto Alegre, Brazil.*

**Abstract :** Oscillatory shear on two-dimensional monodisperse liquid foams was performed. We show that the effect of the oscillatory shear is to cause the migration of bubbles which size is greater than that of a typical bubble of the foam. These so-called 'flaws' move towards the periphery of the foam in a non random motion, thus realizing size segregation in a system which is by construction gravity insensitive. We also show that elongated cavities in the foam could be relaxed towards a more isotropic form with oscillatory shear, and we discuss the pertinent parameters of this relaxation.

PACS. 05.65.+b – Self-organized systems.

PACS. 83.80.Iz – Emulsions and foams.

Keywords : rheology of two-dimensional foams, individual motion, size segregation, cell sorting.

## Introduction :

Versatility of liquid foam has not yet been exhausted. Its rich rheological behaviour, due to its visco-elasto-plastic nature, is still to be fully understood [1,2]. On the point of vue of physics, the state of a liquid foam is almost resumed by its geometrical and topological features. This, added to the fact that they are easy to produce and handle and, for two-dimensional (2D) ones (*i.e.* reduced to a single layer of bubbles), easy to image, make it a quite efficient model for other complex systems such as amorphous or granular materials[1,3], or even biological ones[4-6]. Among the questions that interest both physicists and biologists, cell motion into a cell aggregate is a complex one : of course, biochemical events mainly drive aimed displacements, but purely physical mechanisms do intervene too. The role of adhesion and interfacial energy between comparable / different cells was evocated long ago, and is still fertile[7-10]. The role of the size of the cells involved in a motion has received less attention. This very basic feature is nevertheless easily controllable in foam experiments and could be of great importance in relative motions, as suggested by counter-intuitive effects observed in 2D foams[11]. The important point when studying motions in a liquid foam is that the energy needed for bubble reorganization is much higher than the thermal activation energy. Simply heating the foam would not be compatible with its stability requirements. However, relaxation from a quenched state may be achieved through the successive deformation cycles applied during oscillatory shear[12], which effect is particularly easy to follow through image analysis on a millimetric two-dimensional foam. Couette continuous shearing of a 2D foam has been intensively studied recently[3,13,14], but the connexe geometry of the 2D shear cell described in the following section, and the oscillatory aspect, are more reminiscent of what happens in a deforming cell cluster.

In the following experiments we examine the fate of a single bubble in a two-dimensional monodisperse foam. Two types of bubbles were investigated. Firstly, bubbles of area a few times the one of the surrounding monodisperse bubbles. These will be called 'flaws'. Secondly, bubbles of about 2 orders of magnitude greater area than the surrounding bubbles. These are far more deformable and more accurately described by the term 'cavities. This paper presents

---

[1*,] Corresponding author : Catherine.Quilliet@ujf-grenoble.fr

the first results both on the migration of a 'flaw' and on the relaxation in shape of an anisotropic 'cavity', under oscillatory shear.

**Experimental part :**

The 2D foams studied here are formed the same way as Fortes and Rosa[15]. They consist of a single layer of bubbles sandwiched between (soapy) water and a plate glass cover. Besides previously mentioned features proper to two-dimensional foams, the use of a horizontal system avoids heterogeneity problems due to drainage[1]. Two more advantages of this so-called "lisboetan configuration" are that (i) the equilibrium of the soap solution is very quickly reached between the reservoir solution lying under the foam and the foam itself, and (ii) the foam is mechanically accessible from the solution under it, allowing manipulations much more easily than when confined between two plates.
The soap solution is 15% volume Dreft[16] and 5% weight sugar into water. Sugar is added to lengthen foam life by preventing evaporation ; very probably its presence is a superfluous precaution in the closed set-up presented here, as explained in the next paragraph. The foam is formed by slowly blowing nitrogen bubbles into the solution. When the gas flow rate is slow enough, the size of the bubbles is dependent only on the pipe geometry and on the physico-chemical properties of the solution[17]. We used a flow rate around 10 mL/min, which allowed us to obtain a quite monodisperse foam : the standard deviation $\sigma_{2D}$ of the distribution of bubble areas in the 2D foam is of order 0.06 times the average bubble area. One characterizes the foam by its *typical bubble size*, which is the square root of the average bubble area.

Four barriers confined the foam (2 mm thick) in a rectangle between the glass cover and the water surface. Two of the barriers were made from an elastic rubber, that allowed the confinement area to be continuously deformed from a rectangle (fig. 1a) to a parallelogram <u>of same area</u> (fig. 1b). This allows the shear of the foam in its own plane, without compressing it. What we call in the following a *shear cycle* consists in shear up to some angle $\theta_{max}$ of the parallelogram, then reversing the tilt up to the inverse angle $-\theta_{max}$, and finally coming back to the initial rectangular position of the barriers (fig. 1 illustrates the whole cycle). The shearing is done continuously, and slowly enough to be considered as quasistatic (shear rate $\dot{\gamma} \sim 0.1\ s^{-1}$). It is interesting to note that the barriers also prevent the direct contact of the foam borders with air. Contact with room atmosphere usually weakens boundary bubbles due to pressure variations and evaporation on their free faces. Protected from the first cause of bubbles breakage, the foams are thus stable on the time scale of the experiment and over, and do not show coarsening either[18].
The 'flaws' were made by simply blowing a single (bigger) bubble near the center of the rectangle occupied by the 2D foam. The cavities were realized by first plating a non-revolution hollow cylindrical piece (laboratory version of biscuit cutters) under the glass cover, then blowing the foam outside of it. The 2D foam spreads all around the cylinder, and the cavity instantaneously but not much relaxes from the cutter shape when this latter is removed. The stability of the foam in this set-up is such that the aspect of the whole (foam + shaped cavity) can stay unmodified for at least a dozen of hours.

The observations are made from above the glass plate, using a CCD camera. A neon ring, of diameter 40cm, is used to illuminate the foam from several centimeters above the glass plate. This lighting provides images with excellent contrast, that are then digitalized and analyzed with the software Scion Image[19].

**Results and discussion :**

1) Migration

A 'flaw' was inserted near the middle of the foam. The foam was then sheared, as described above, using an angle $\theta_{max}$ = 29º (*i.e.* a shear strain $\gamma$ = 0.55, fig. 1b). In order to measure the displacement of the flaw, a picture was taken when the barriers containing the foam came back to their initial rectangular position. The distance that the flaw had moved from its initial position was plotted versus the number n of shear cycles.

Figure 2 displays typical migration curves for a flaw (closed symbols). The plots show an initial rapid increase in distance from the start position, following by a slowing down of the motion as the flaw approaches a border of the parallelogram (no preference was observed concerning the border "choosen"). Once the flaw was within 2-3 rows from a wall, it did not return to the center, even after numerous shear cycles (up to n=35). This clear attraction of the flaw towards the periphery recalls outwards segregation of bigger bubbles already (i) observed in 3D foams under continuous plate-plate shear[20], and (ii) obtained from numerical simulations performed on 2D foams under uniaxial compression[21]. At the moment we do not have an explanation for such behavior, but the sticking at some bubbles diameters from the periphery is consistent with Abd el Kader's and Earnshaw's observation, in a similar geometry, that motion is mainly concentrated in the central regions[18].

In the initial regime of rapid increase, the curve appears somewhat irregular due to the non-continuous nature of the medium in which the flaws do move by jumps. Nevertheless, the global motion is almost linear up to n~10, and could hardly be approximated by a square root of n, even for the first cycles where the concavity is more upwards. Hence the flaw does not appear to have a random walk, which implies that in these foams the oscillatory shear would not have the mere effect of thermal activation : this goes in the same sense than what was shown by Viasnoff et al in dense colloidal suspensions[12]. In order to confirm the linear dependence of the distance from initial position versus shear cycles, it would be necessary to perform comparable experiments with a larger set-up. This would have the effect of smoothing the data by reducing the relative importance of jumps.

In a second set of quantitative measurements, we took advantage of the relaxation experiments exposed hereafter to get an insight into the migration of much bigger bubbles, *i.e.* cavities. Motion of the barycentre of the cavity is displayed on figure 2 (open symbols) for four runs. Three of them are relevant of the usual behaviour of cavities, which stay near the center even for a large number of shear cycles. In a small proportion of the experiments however, the cavities migrate to the periphery just as flaws (fig. 2, open circles). We did not identify up to now the key parameter for the occurrence of these cases, but we finally could statistically distinguish between the features of flaws and cavities motion, by extracting the distance by which they move in one shear cycle, and normalizing it by the typical size of foam bubbles. As the mere plot of this quantity does not show evolution with the number of shear cycles, we grouped all the data into a double histogram (fig. 3). The difference is clear : for the flaws, the distribution of the normalized jumps is wide and displaced towards high values (maximum between 1 and 1.2). The cavities have a maximum jump probability between 0.2 and 0.4 times the typical bubble size, and a narrower distribution. The reduced mobility of cavities is easy to understand, because they are bigger than the flaws their movement involves more changes in the foam bubbles conformation. However this local size effect is not sufficient to explain the discrepancy between the 'macroscopic' resulting motions showed on figure 2 : as the curves for flaws and for cavities can obviously not be proportional, which should be the case for randomly oriented jumps, the explanation for their differences (orientation of jumps by the shear ?) has to be search within the nontrivial rheology of 2D foams.

2)Anisotropy relaxation

Inspired by the beautiful experiments of Rieu[22] on the relaxation of two-dimensional anisotropic cell aggregates, we decided to study the relaxation of elongated islands of

2D foams. But, as our way of bringing energy, *i.e.* shearing, is not compatible with the finite extension of a cluster, we looked at the complementary geometry, which is an elongated cavity (obtained with a rectangular 15x45 mm cutter) into a monodisperse foam.

As shown on fig. 4, these experiments allow a quite efficient relaxation of the cavity. It is interesting to remark that the cavity quite often relaxes in a facetted shape, confirming in an inverse geometry the predictions by Cox and Graner [23] about the crystal behaviour of bubble clusters. Quantitative measurements were obtained for oscillatory shear at $\theta_{max}$ =29º (the maximum angle allowed by our set-up), by calculating the cavity deformation parameter $\varepsilon = 1 - b/a$, where b and a are respectively the short and long axis of the ellipse that best fits the cavity perimeter (both values provided by Scion Image).

The evolution of $\varepsilon$ with the number n of shear cycles shows a decrease followed by a plateauing at some final value $\varepsilon_f > 0$ (figure 5), as was observed in numerical simulations[18]. In most of the experiments, the decrease takes place on less than a decade in $\varepsilon$, but in this extend it is consistent with the exponential decay shown in two-dimensional cellular aggregates[19].

From semi-log plots as in fig. 5, we extracted the decay parameter $n_0$, such as $\varepsilon = \varepsilon_{initial} \exp(-n/n_0)$. All our experiments over, $n_0$ varied between 0.38 (relaxation towards a shape coarsely round to the eyes in one shear cycle) and 6.62. Contrary to what could be expected, no correlation between the decay parameter $n_0$ and the foam typical size (ranging here from 3.3 to 6.6 mm) nor with the foam thickness (between 1 and 3 mm) could be put into evidence. Furthermore the decay values were hardly reproducible between two experiments of comparable foam typical size and thickness, and cavity shape. After numerous trials, we suspect the degree of order to be of higher influence on the decay parameter than the other parameters. However an important systematic work has to be done to distinguish between the roles of mere polydispersity (fixed for a given foam) and its ordering which can be followed by calculating the width $\mu_2$ of the distribution of the bubbles number of sides.

For what concerns the plateau value $\varepsilon_f$, it is clearly dependent on the shear angle $\theta_{max}$ (fig. 6). The larger the shear angle, the better the relaxation (*i.e.* towards a more isotropic shape). But for the widest angle the important bubble rearrangements at every shear cycle, leads to quite scattered values of the deformation parameter. This seems to indicate that oscillatory shear ceases to order the medium for high shear strains.

**Conclusion :**

These experiments showed that promising results could be achieved with a very simple shearing set-up. Relaxation of anisotropic cavities is very efficiently realized in a monodisperse foam, up to some plateauing value of the deformation, which depends on the shear angle. The decrease of the deformation rate shows no correlation with the average geometrical features of the foam, but ordering of the latter could be of some importance. The migration experiments indicate on the one hand a non-random motion of 'flaws', showing the difference between oscillatory shear and thermal energy fluctuations. On the other hand, both attraction of flaws near the periphery and possible incidence of their size on the motion make expect the possibility of observing collective size segregation in 2D foams.

**Acknowledgements :** Authors thank CAPES-Cofecub program for financial support, which made this franco-brasilian collaboration effective, and C.Q. thanks Andrew Campbell for his help in english redaction.

Figure captions :

Figure 1 : Stages of a shear cycle in a migration experiment, observed from above. The lower boundary is fixed, whilst the upper boundary fixed to the cover glass is allowed to slide. This causes the initially perpendicular boundaries to incline and elongate, transforming the rectangle in a parallelogram of same area. (a) Initial state (or cycle n=0). Typical size of foam bubbles is 3.9 mm ; area ratio 2.47 between flaw and foam bubbles. (b) Shear forth ; here the shear angle $\theta$ is the maximum value $\theta_{max}$ attained in the shear cycle ($\theta_{max}$ =29°). (c) Maximum strain back. (d) Back to initial position, end of first shear cycle (n=1). Length of fixed barrier : 103 mm.

Figure 2 : Closed symbols : distance between current and initial position of the flaw, as a function of the number n of shear cycles ($\theta_{max}$ = 29°). Diamonds : size of foam bubbles 3.9 mm, area ratio 2.5 between flaw and foam bubbles. Flaw within 3 rows from the periphery at n=13. Squares : size of foam bubbles 3.8 mm, area ratio 2.5. Flaw within 3 rows from the periphery at n=10. The point at n = 0.5 is related to the first passage through initial position after the first shear forth (intermediate picture between fig. 1b and 1c). Triangles : size of foam bubbles 4,0 mm, area ratio 2. Open symbols : Diamonds, squares, circles, and triangles represent, respectively, experiments with cavities of area ratio 43, 45, 63, 72 and size of foam bubbles 4.4, 3.9, 3.7 and 3.3 mm.

Figure 3 : Histogram of the distribution of normalized jumps, i.e. the distance by which the flaw/cavity moves in one shear cycle, divided by the typical size of foam bubbles. Grey : flaws, 4 experiments with flaw/foam bubbles area ratio between 2 and 7. Black : cavities, 4 experiments with area ratio between 43 and 72.

Figure 4 : Relaxation experiment on an elongated cavity in a monodisperse foam (>800 bubbles of characteristic size 3.3 mm). (a) Initial state after removing the rectangular 15x45mm *biscuit cutter* that shapes the cavity : deformation parameter $\varepsilon$ = 0.60. (b) After n=7.5 shear cycles ($\theta_{max}$ = 29°) : $\varepsilon$ = 0.16. The quite high ordering of the foam explains the facetted aspect of the rounded cavity. Length of "horizontal" boundaries: 103 mm.

Figure 5 : Deformation parameter $\varepsilon$ versus number n of shear cycles (semi-log plot), for the experiment shown fig. 4 (maximum shear angle $\theta_{max}$ = 29°). Linear plain curve : $\varepsilon = \varepsilon_{initial} \exp(-n/n_0)$, with $n_0$=2.89. The plateau value (indicated by the dotted line) is denoted as $\varepsilon_f$. Note that the scattering in this part of the relaxation curve is enhanced by the semi-log representation.

Figure 6 : Multiple relaxation experiment : deformation parameter $\varepsilon$ versus number n of shear cycles (linear plot) with increasing maximum shear angle $\theta_{max}$ by steps (up to n=5 : $\theta_{max}$=9° ; then $\theta_{max}$=18° up to n=9.5 ; then $\theta_{max}$=29°). Foam of size 5.3 mm is displayed on the picture in its initial state (length of "horizontal" boundaries : 103 mm). Upper insert : plateau values $\varepsilon_f$ (obtained from main curve) versus maximum shear angles $\theta_{max}$. The dotted line is a guide for the eyes.

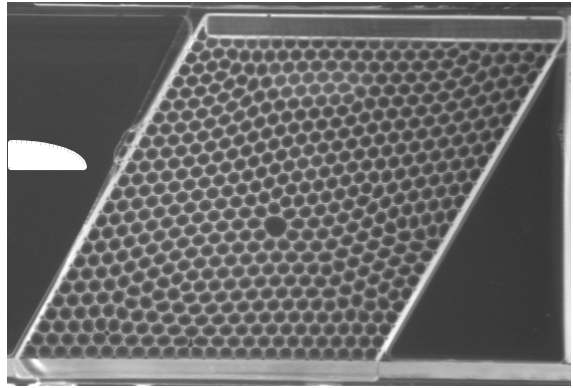

Figures 1a, 1b

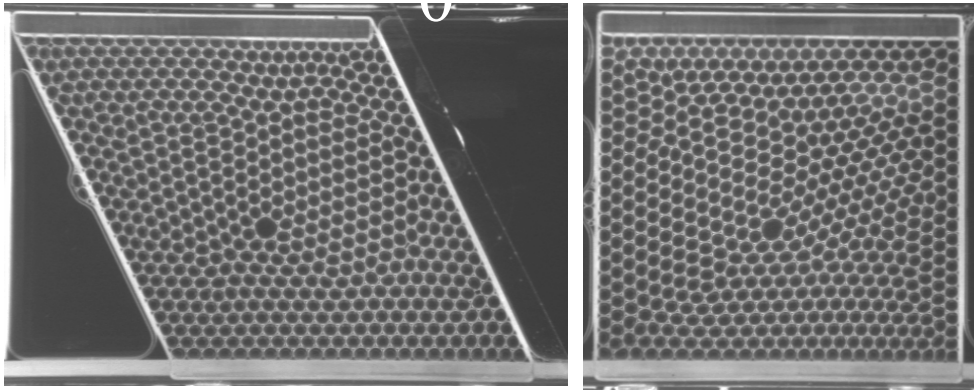

Figures 1c, 1d

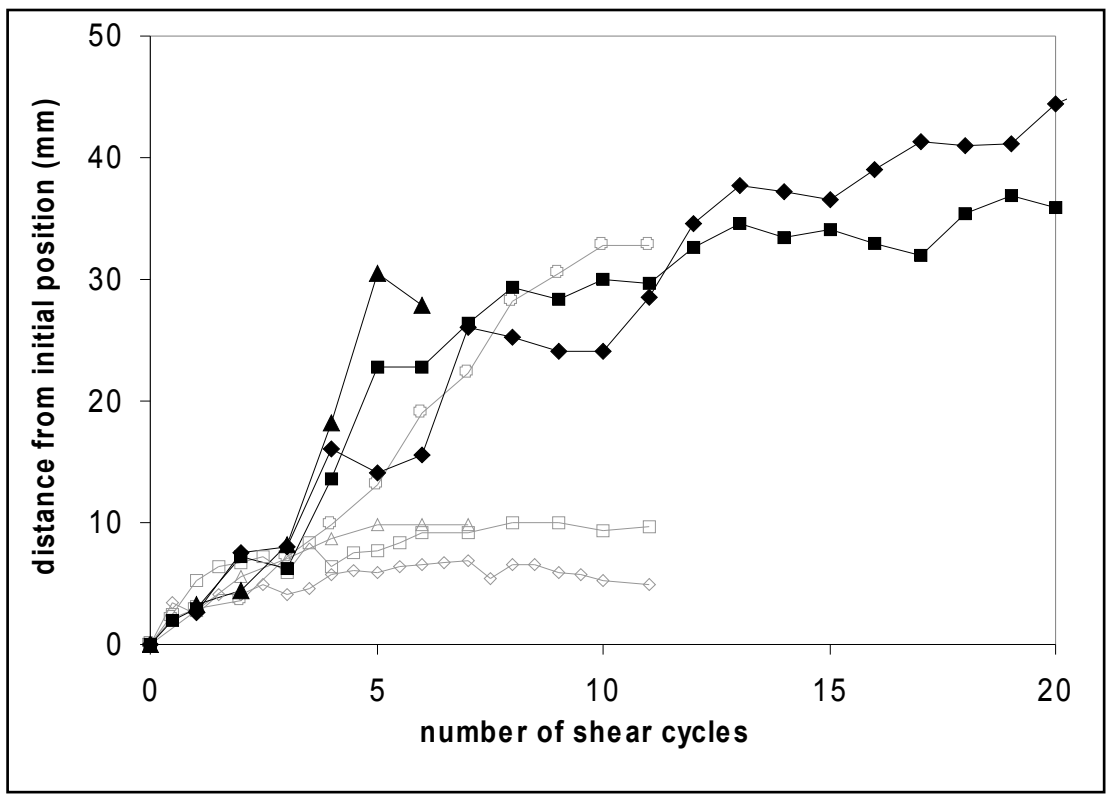

Figure 2

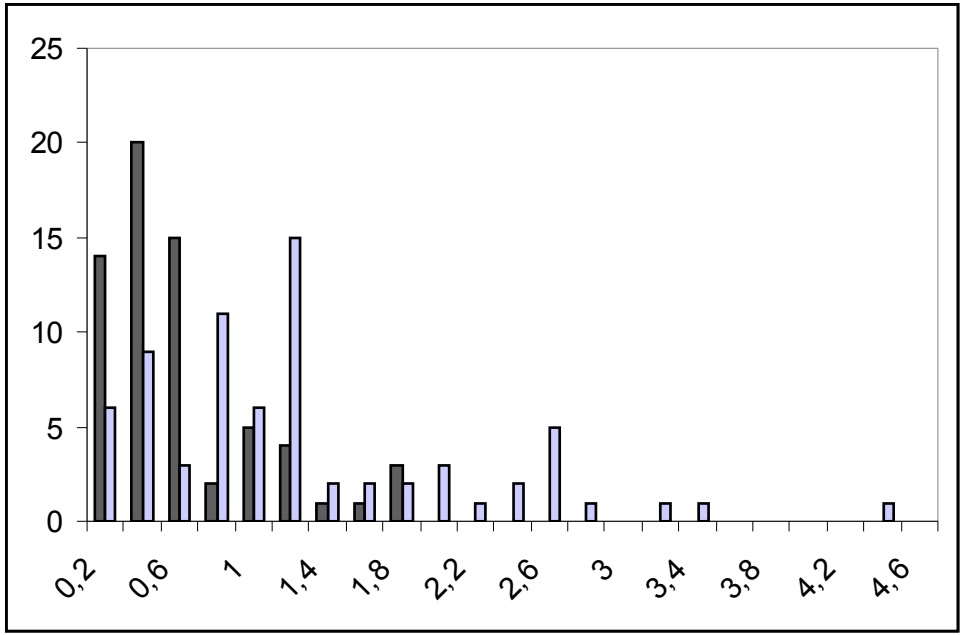

Figure 3

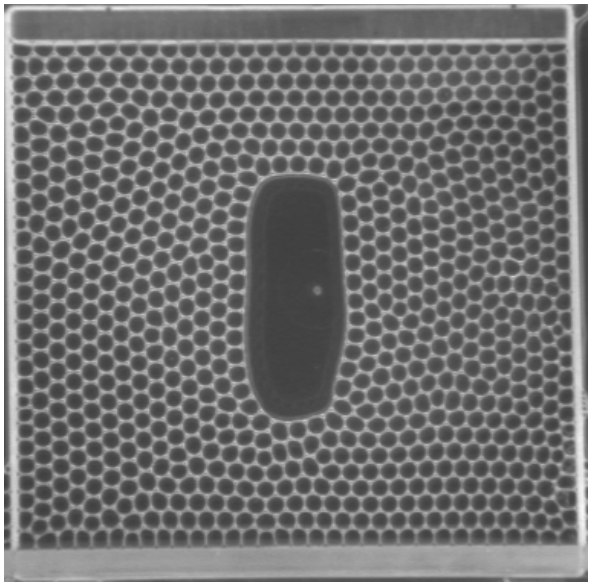
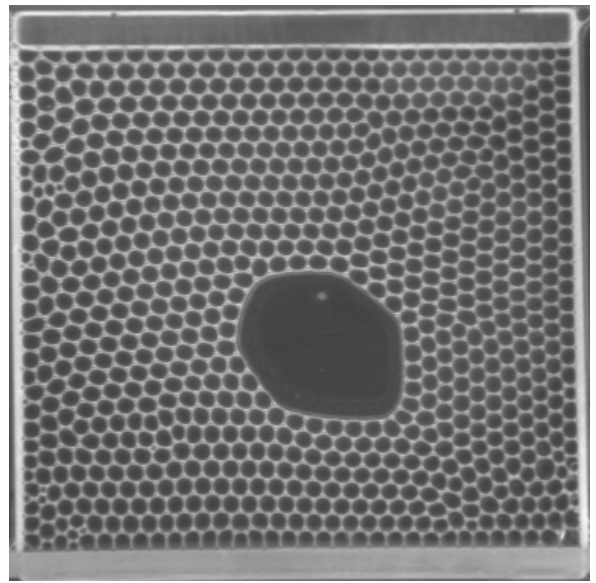

Figures 4a, 4b

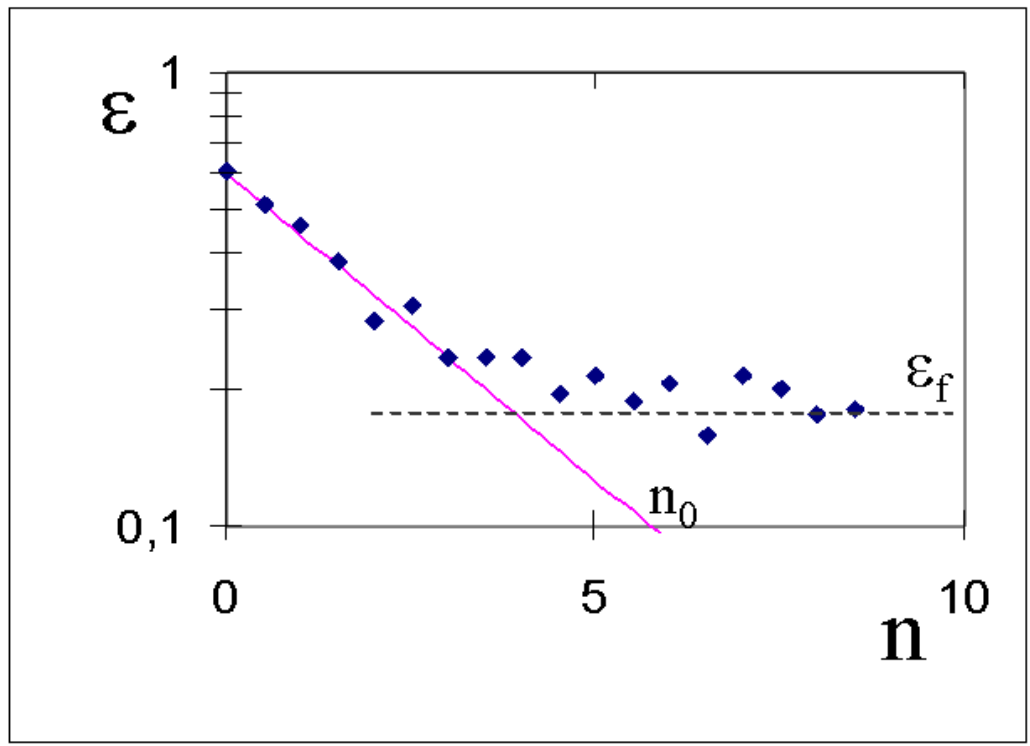

Figure 5

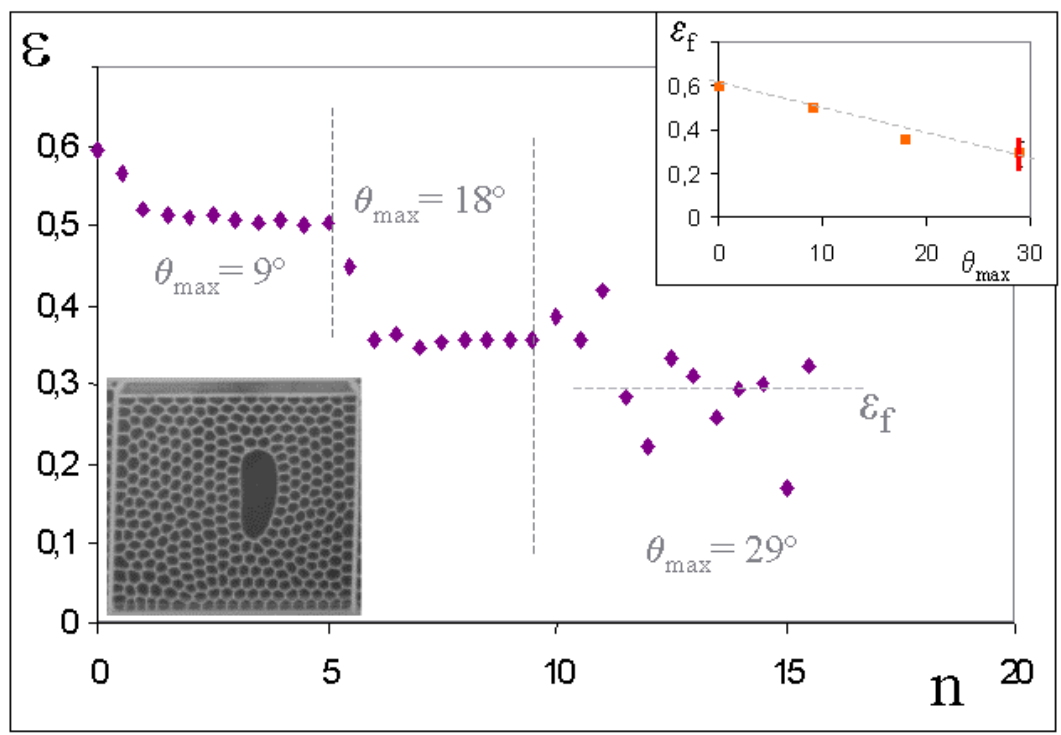

Figure 6